\UseRawInputEncoding
\documentclass[%
reprint,
superscriptaddress,
nofootinbib,
 amsmath,amssymb,
 aps,
]{revtex4-1}
\usepackage{graphicx}% Include figure files
\usepackage{subfigure}
\usepackage{dcolumn}% Align table columns on decimal point
\usepackage{bm}% bold math

\usepackage[%
bookmarksnumbered,
bookmarksopen,
colorlinks,
citecolor=blue,
linkcolor=blue,
]{hyperref} % add hypertext capabilities
\def\es0{$E_{sym}(\rho_0)$~}
\def\us0{$U_{sym}^{\infty}(\rho_{0})$}
\begin{document}

%\preprint{APS/123-QED}

\title{Kaon production in the HADES experiment in Au+Au collisions at $\sqrt{s_{\rm NN}}=2.4$~GeV}

\author{Gao-Feng Wei}\email{Corresponding author: wei.gaofeng@gznu.edu.cn}
\affiliation{School of Physics and Electronic Science, Guizhou Normal University, Guiyang 550025, China}
\affiliation{Guizhou Provincial Key Laboratory of Radio Astronomy and Data Processing, Guizhou Normal University, Guiyang 550025, China}
\author{Yu-Liang Zhao}
\affiliation{School of Physics and Electronic Science, Guizhou Normal University, Guiyang 550025, China}

\begin{abstract}

Within an isospin- and momentum-dependent transport model by including the kaon reaction channels, we study the kaon prodution in heavy-ion collisions (HICs) at SIS (Darmstadt Schwerionen Synchrotron, GSI) energies. Based on simulations of a centrality of 0-40\% Au + Au collision at $\sqrt{s_{NN}}=2.4$~GeV, a typical reaction that has been carried out by the HADES Collaboration, we confirm that the medium modification of kaon masses plays a vital role in studying the kaon productions in HICs, and is also unavoidable for the successful interpretation of the HADES data on kaon rapidity distributions and transverse mass spectra. Moreover, it is shown that the directed flows of kaons are affected significantly by the kaon potential and slightly affected by the medium modification of kaon masses. Also, the rapidity-dependent inverse slope parameter $T_{B}$ of the kaon transverse mass spectra is shown to be affected considerably by both the kaon potential and medium modification of kaon masses. However, through checking the simulations of related reactions in FOPI and/or KaoS experiments, some of these regular effects do not seem to be obvious and appear to be the reaction system and/or beam energy dependent. Nevertheless, it can be confirmed that the medium modification of kaon masses is favored by observations from the inverse slope parameter $T_{B}$ and transverse mass spectra of kaons in both HADES Au + Au collisions at $\sqrt{s_{NN}}=2.4$~GeV and FOPI Ni + Ni collisions at 1.93\textit{A} GeV. Therefore, measurements of the inverse slope parameter $T_{B}$ of kaon transverse mass spectra and the kaon directed flows in HADES Au + Au collisions would be great benefit to detection of the kaon potential and the corresponding medium effects on kaon masses.
\end{abstract}

\maketitle

%\tableofcontents

\section{introduction}\label{introduction}

The study of hot and dense matter in HICs is known to be closely related to the determination of equation of state (EoS) of nuclear matter and evolution of neutron stars as well as the properties of hadrons in medium~\cite{Hor14,Heb15,Lat16,Bal16,Oer17,Hana17}. Of particular interest, especially after Kaplan and Nelson proposed the likelihood of kaon condensation in dense matter~\cite{Kap86}, the kaon production in HICs has been paid much attention in both theories~\cite{Aich85,LiGQ94,LiBA94,Song99,Fuch01,Sturm01,Hart06,Hart12,Scha94,Brown94,LiGQ97,LiGQ97b,Cassing97,AB05,Gasik16,Song21,Song22} and experiments (e.g., KaoS~\cite{KaoS94,KaoS97,KaoS98,KaoS05,KaoS07}, FOPI~\cite{FOPI-Kaon97,FOPI-Kaon00,FOPI-Kaon07,FOPI-Kaon09,FOPI-Kaon14} and HADES~\cite{HADES09,HADES10,HADES11,HADES14,HADES18,HADES19,HADES20} experiments), since the kaon production in HICs especially at SIS energies of 1-2\textit{A} GeV is shown to be a more promising probe for the study of both EoS of nuclear matter~\cite{Aich85,Fuch01,Hart12} and medium modification on hadron properties due to chiral symmetry restoration~\cite{Scha94,Brown94,LiGQ97,LiGQ97b,Cassing97,AB05,Gasik16,Song21}.  

So far, hadron properties and interactions at high baryon densities cannot be addressed directly by \textit{ab initio} QCD calculations, and thus have to be studied in HICs~\cite{HADES18}. Specifically, for the kaon production in HICs, both the kaon potential or dispersion relation and medium modification of kaon masses might be the two key factors. For the former, through comparing the data on flows and spectra of kaons~\cite{KaoS07,FOPI-Kaon09,FOPI-Kaon14,HADES10} to transport model calculations, most studies favor qualitatively a repulsive kaon potential~\cite{Cassing97,Pal00,Hart12,Metag17,LiGQ95,LiBA96,LiBA99}, however, with different strengths even different forms, see, e.g., Refs.~\cite{LiGQ95,LiBA96,LiBA99}. While for the latter, it has been pointed out that medium modification of kaon masses could increase in-medium kaon effective mass and thus the threshold of kaon production in HICs, e.g., in Refs.~\cite{Cassing97,Gasik16,LiGQ97}. However, it seems to no firm conclusions to be reached. To this situation, it is naturally necessary to examine effects of both different kaon potential or dispersion relation and the corresponding medium modification of kaon masses on kaon production in HICs at SIS energies. Stimulated by the recently reported HADES data on kaons~\cite{Song21,HADES18,HADES19}, we conduct this study and thus expect to shed light on the mechanism of kaon production in HICs at SIS energies.
On the other hand, based on the HADES nucleon and light cluster data~\cite{HADES20,ECT}, we have studied and reproduced reasonably the nucleon and deuteron rapidity distributions as well as collective flows with an isospin- and momentum-dependent Boltzmann-Uehling-Uhlenbeck (IBUU) transport model, see, Ref.~\cite{Wei23} for the details.
Therefore, we also want to know whether our model could reproduce reasonably the HADES kaon data. To this end, within two scenarios for the kaon potential or dispersion relation as well as the corresponding medium modification on kaon masses, i.e., empirical scattering length and chiral Lagrangian scenarios~\cite{Pal00,Metag17,LiGQ95,LiBA96,LiBA99}, we perform a centrality of 0-40\% Au + Au collision at $\sqrt{s_{NN}}=2.4$~GeV. Moreover, we simulate the related reactions in FOPI and/or KaoS experiments, and compare our results with the corresponding data. It is shown that the medium modification of kaon masses is favored by both HADES Au + Au collisions at $\sqrt{s_{NN}}=2.4$~GeV~\cite{Song21,HADES18} and FOPI Ni + Ni collisions at 1.93\textit{A} GeV~\cite{FOPI-Kaon97}. It therefore can be confirmed that the medium modification of kaon masses plays a vital role in studying the kaon productions in HICs at SIS energies.

\section{The Model}\label{Model}
This study is carried out within an IBUU transport model similar to that used in our previous study in Ref.~\cite{Wei23}.
Compared with the original IBUU model~\cite{IBUU1,IBUU2}, the present version of IBUU model includes mainly two aspects of updates. 
On the one hand, for more delicate treatment of the in-medium many-body force effects as pointed out in Refs.~\cite{Xu10,Chen14}, we adopt a separate density-dependent scenario~\cite{Wei20}. Also, to meet the best knowledge of symmetry energy at the saturation density $\rho_{0}$ and $2\rho_{0}/3$ one has obtained, e.g., in Refs.~\cite{Cozma18,Wang13,Wang18,Ess21,Est21,Ree21}, we introduce a parameter $z$~\cite{Wei22} as in Ref.~\cite{Xu15} to mimic the symmetry energy at the two densities. 
Specifically, the isospin and momentum-dependent nuclear interaction (MDI) used is expressed as:
\begin{eqnarray}
U(\rho,\delta ,\vec{p},\tau ) &=&A_{u}\frac{\rho _{-\tau }}{\rho _{0}}%
+A_{l}\frac{\rho _{\tau }}{\rho _{0}}+\frac{B}{2}{\Big(}\frac{2\rho_{\tau} }{\rho _{0}}{\Big)}^{\sigma }(1-x)  \notag \\
&+&\frac{2B}{%
	\sigma +1}{\Big(}\frac{\rho}{\rho _{0}}{\Big)}^{\sigma }(1+x)\frac{\rho_{-\tau}}{\rho}{\big[}1+(\sigma-1)\frac{\rho_{\tau}}{\rho}{\big]}
\notag \\
&+&\frac{2C_{l }}{\rho _{0}}\int d^{3}p^{\prime }\frac{f_{\tau }(%
	\vec{p}^{\prime })}{1+(\vec{p}-\vec{p}^{\prime })^{2}/\Lambda ^{2}}
\notag \\
&+&\frac{2C_{u }}{\rho _{0}}\int d^{3}p^{\prime }\frac{f_{-\tau }(%
	\vec{p}^{\prime })}{1+(\vec{p}-\vec{p}^{\prime })^{2}/\Lambda ^{2}},
\label{IMDIU}
\end{eqnarray}%
where $\tau=1$ for neutrons and $-1$ for protons, and $A_{u}$, $A_{l}$, $C_{u}(\equiv C_{\tau,-\tau})$ and $C_{l}(\equiv C_{\tau,\tau})$ are expressed as
\begin{eqnarray*}
A_{l}&=&A_{l0}+U_{sym}^{\infty}(\rho_{0}) - \frac{2B}{\sigma+1}\notag \\
&\times&\Big{[}\frac{(1-x)}{4}\sigma(\sigma+1)-\frac{1+x}{2}\Big{]},  \\
A_{u}&=&A_{u0}-U_{sym}^{\infty}(\rho_{0}) + \frac{2B}{\sigma+1}\notag \\
&\times&\Big{[}\frac{(1-x)}{4}\sigma(\sigma+1)-\frac{1+x}{2}\Big{]},\\
C_{l}&=&C_{l0}-2\big{(}U_{sym}^{\infty}(\rho_{0})-2z\big{)}\frac{p_{f0}^{2}}{\Lambda^{2}\ln \big{[}(4p_{f0}^{2}+\Lambda^{2})/\Lambda^{2}\big{]}},\\
C_{u}&=&C_{u0}+2\big{(}U_{sym}^{\infty}(\rho_{0})-2z\big{)}\frac{p_{f0}^{2}}{\Lambda^{2}\ln \big{[}(4p_{f0}^{2}+\Lambda^{2})/\Lambda^{2}\big{]}}.
\end{eqnarray*}
%%%
The eight parameters embedded in the above expressions, i.e., $A_{l0}$, $A_{u0}$, $B$, $\sigma$, $C_{l0}$, $C_{u0}$, $\Lambda$ and $z$, are determined by fitting eight experimental and/or empirical constraints on properties of nuclear matter at $\rho_{0}=0.16$~fm$^{-3}$. Among them, the values of first seven parameters are $A_{l0}=A_{u0}=-66.963$~MeV, $B=141.963$~MeV, $C_{l0}=-60.486$~MeV, $C_{u0}=-99.702$~MeV, $\sigma=1.2652$, and $\Lambda=2.424p_{f0}$, where $p_{f0}$ is the nucleon Fermi momentum in symmetric nuclear matter (SNM) at $\rho_{0}$. The eighth parameter $z$ is associated with the symmetry energy parameter $x$ that is used to mimic the slope value $L\equiv{3\rho({dE_{sym}}/d\rho})$ of symmetry energy at $\rho_{0}$. Because we aim in this study to examine kaon production in HICs, we therefore use a certain value of $L=61.95$~MeV and the corresponding $x$ and $z$ parameters are $x=0.2$ and $z=0.326$~MeV. Other physical quantities are also fixed from above used parameters including the binding energy $-16$~MeV, the pressure $P_{0}=0$~MeV/fm$^{3}$, the incompressibility $K_{0}=230$~MeV for SNM at $\rho_{0}$, the isoscalar effective mass $m^{*}_{s}=0.7m$, the isoscalar potential at infinitely large nucleon momentum $U^{\infty}_{0}(\rho_{0})=75$~MeV, the isovector potential at infinitely large nucleon momentum $U^{\infty}_{sym}(\rho_{0})=-100$~MeV as well as the symmetry energy $E_{sym}(\rho_{0})=32.5+0.326$~MeV and $E_{sym}(2\rho_{0}/3)=25.1$~MeV. For the details, we refer readers to see Refs.~\cite{Wei23,Wei20,Wei22}. 

On the other hand, to study kaon production in HICs, we further develop our model to include kaon production and annihilation as well as elastic channels as will be discussed later in combination with the results in Sec.~\ref{Results and Discussions}. Moreover, to examine effects of the kaon potentials as well as the corresponding medium modification of kaon masses in HICs, we adopt two commonly used scenarios as pointed out in many literatures as aforementioned, e.g., in Refs.~\cite{LiBA96,LiBA99,LiGQ95,Pal00,Metag17}. The first is based on the kaon dispersion relation determined from kaon nucleon scattering length using the impulse approximation, i.e., 
\begin{equation}\label{ESL-relation}
\omega_{K}=\Big{[}m_{K}^{2}+{{\bf k}}^{2}-4\pi\alpha_{\small{KN}}\Big{(}1+\frac{m_{K}}{m_{N}}\Big{)}\rho\Big{]}^{1/2},
\end{equation}
where $m_{K}$ and $m_{N}$ are the kaon and nucleon masses in free space, ${\bf {k}}$ is the kaon momenta, $\rho$ is the baryon densities, and $\alpha_{KN}\approx{-0.255}$~fm is the kaon nucleon scattering length, which leads to a repulsive kaon potential of 30~MeV at $\rho_{0}$. Another is the scalar-vector kaon potential determined from the chiral Lagrangian approach defined as,
\begin{equation}\label{chL-relation}
\omega_{K}=[m_{K}^{2}+{\bf k}^{2}-a_{K}\rho_{S}+(b_{K}\rho)^{2}]^{1/2}+b_{K}\rho,
\end{equation}
where $b_{K}=3/(8f_{\pi}^{2})\approx{0.333}$~GeV~fm$^{3}$, and $\rho_{S}$ is the scalar density. In the mean-field approximation and only the Kaplan-Nelson term is considered, the parameter $a_{K}$ could be expressed as $a_{K}=\Sigma_{KN}/f_{\pi}^{2}$~\cite{Pal00,LiGQ95,LiGQ97}. However, since the value of $\Sigma_{KN}$ has rather large uncertainties, it is therefore phenomenological determination of $a_{K}$ might be a more reasonable method as shown in Refs.~\cite{LiGQ95,LiGQ97,Pal00}. Naturally, the scalar-vector kaon potential determined in Eq.~(\ref{chL-relation}) is nonrelativistic and noncovariant as that in Eq.~(\ref{ESL-relation}). To compare effects of two kaon potentials from above two dispersion relations on kaon production, it is necessary to ensure the values of two kaon potentials at $\rho_{0}$ are identical. With this consideration, we use the value of $a_{K}\approx{0.173}$~GeV$^{2}$~fm$^{3}$ same as in Refs.~\cite{LiGQ95,LiGQ97,Pal00} that also results in a repulsive kaon potential of 30~MeV at $\rho_{0}$. 

\begin{figure}[hbt]
	\includegraphics[width=\columnwidth]{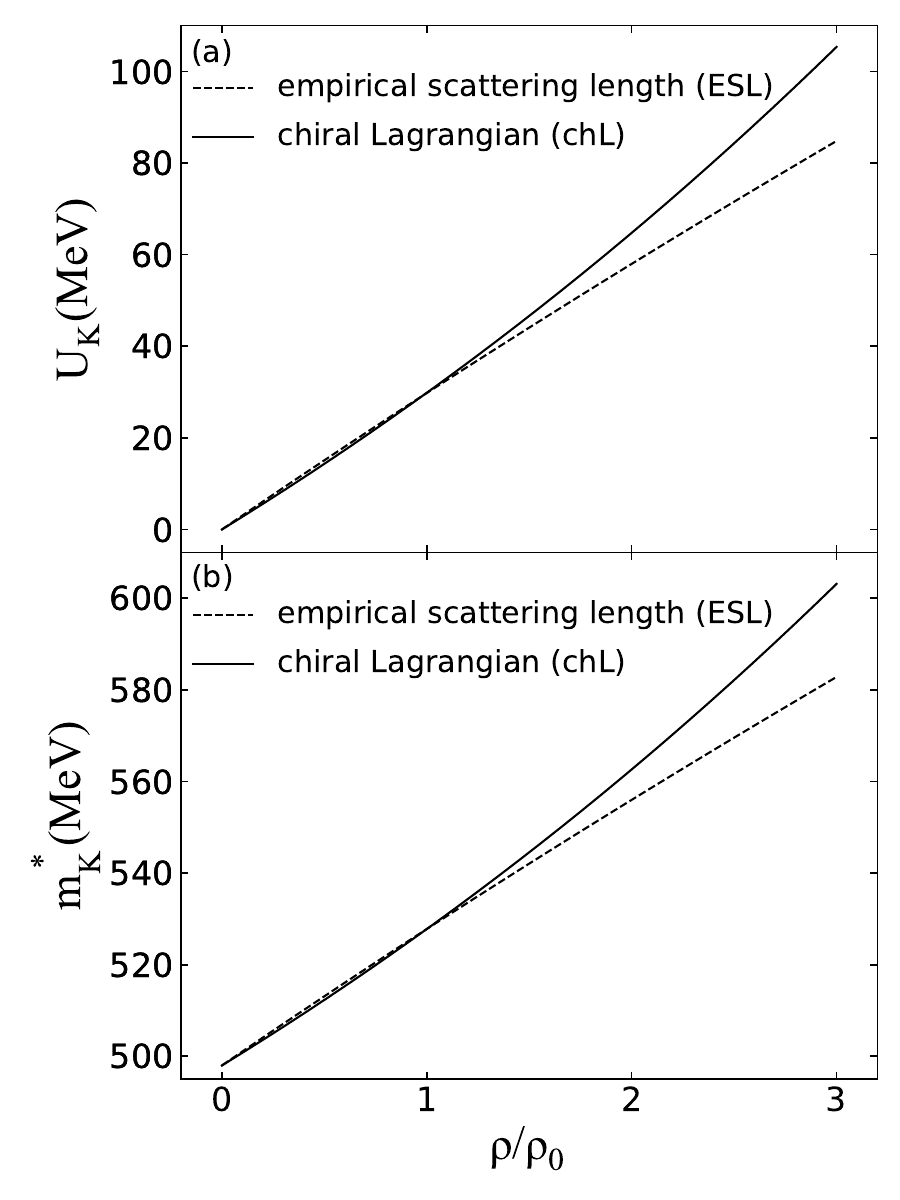}
	\caption{Density dependence of the kaon potentials (a) and the corresponding in-medium effective mass of kaons derived from two dispersion relations, i.e., empirical scattering length (ESL) and chiral Lagrangian (chL).} 
	\label{kaon-pot}
\end{figure}
\begin{figure}[htb]
	\includegraphics[width=\columnwidth]{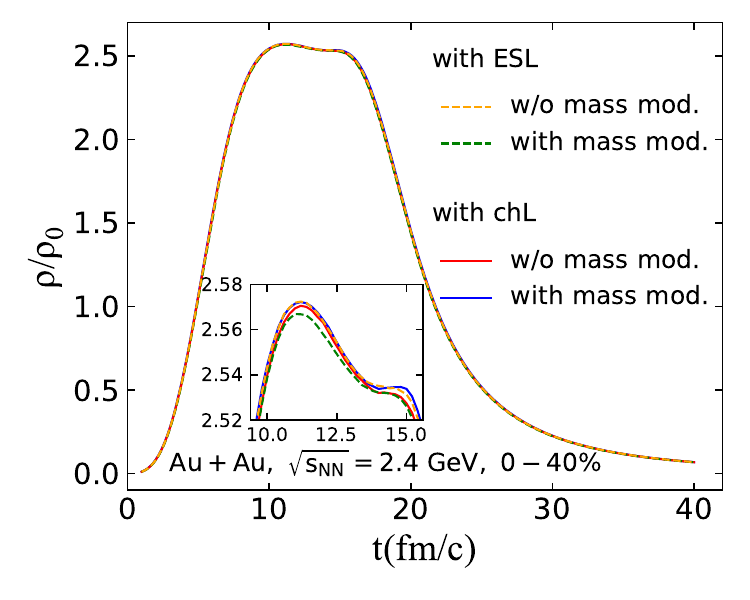}
	\caption{(Color online)Evolution of the central region reduced density $\rho/\rho_{0}$ with and without the mass modification (abbreviated as with mass mod. and w/o mass mod.) in 0-40\% Au+Au collisions at $\sqrt{s_{\rm NN}}=2.4$~GeV. The inset is the local amplification to explicitly show the effects of both kaon potential scenaros and the corresponding medium modification of kaon masses on the compression density, “w/o” stands for “without”.} 
	\label{den}
\end{figure}
\begin{figure*}[thb]
	\includegraphics[width=\textwidth]{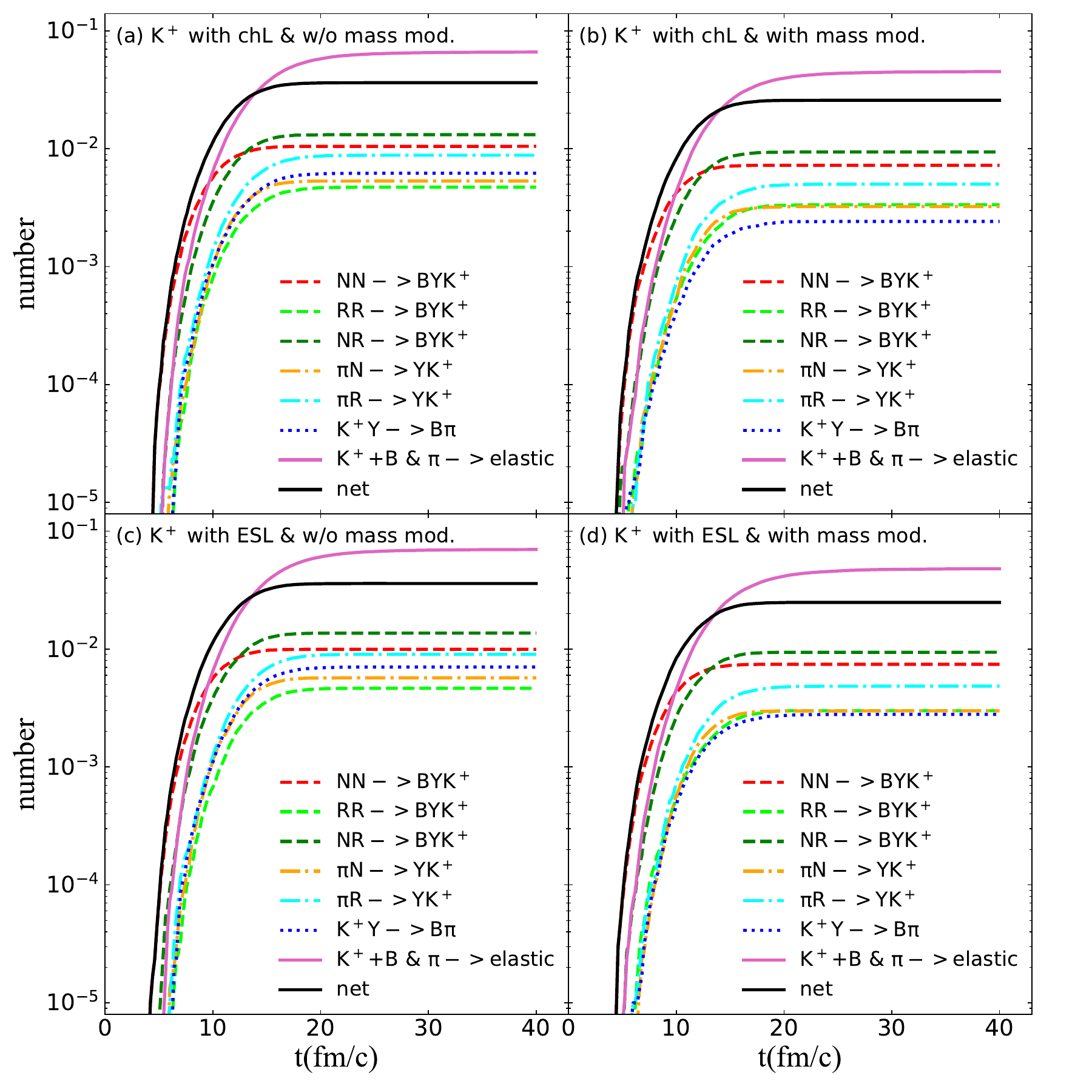}
	\caption{(Color online)Evolution of the kaon number from various channels in 0-40\% Au+Au collisions at $\sqrt{s_{\rm NN}}=2.4$~GeV. The symbol $N$ denotes nucleons, $R$ is the $\Delta(1232)$ or $N^{*}(1440)$, $Y$ is the $\Lambda$ or $\Sigma$, and $B$ is the baryon, i.e., nucleons or resonances.}
	\label{num}
\end{figure*}

To help understand how the kaon potentials affect the kaon production in HICs, we show the two kaon potentials derived from
\begin{equation}\label{pot}
U_{K}(\rho,{\bf k})=\omega_{K}-(m_{K}^{2}+{\bf {k}}^{2})^{1/2},
\end{equation}
as a function of reduced density $\rho/\rho_{0}$ at the zero momentum as shown in Fig.~\ref{kaon-pot}(a). It is seen that the two kaon potentials indeed become divergent as the density increases, however, less than 20~MeV even up to three times the saturation density.  Apart from the kaon potential, the kaon effective mass in dense nuclear medium is also a key factor for kaon production in HICs as aforementioned. Therefore, we also consider medium modification of kaon masses in our simulations of the reaction. Specifically, for the empirical scattering length scenario, the kaon effective mass in medium is defined as
\begin{equation}
	m_{K}^{*}=\Big{[}m_{K}^{2}-4\pi\alpha_{\small{KN}}\Big{(}1+\frac{m_{K}}{m_{N}}\Big{)}\rho\Big{]}^{1/2},
\end{equation}
while for the chiral Lagrangian scenario, the kaon effective mass is expressed as,
\begin{equation}
	m_{K}^{*}=[m_{K}^{2}-a_{K}\rho_{S}+(b_{K}\rho)^{2}]^{1/2}+b_{K}\rho.
\end{equation}
Because our model is not fully relativistically covariant, we therefore estimate the scalar density $\rho_{S}$ as in Refs.~\cite{LiBA95,ART}, i.e.,
\begin{eqnarray}\label{scalar-density-cal}
\rho_{S}&=&\sum_{i=N,\Lambda,\Sigma}\int{\frac{d^{3}{\bm p}}{(2\pi)^{3}}}f_{i}({\bm r},{\bm p},t)\frac{m}{e}\notag\\
&+& \sum_{i=\Delta,N^{*}}\int{\frac{d^{3}{\bm p}}{(2\pi)^{3}}}\frac{dm}{2\pi}f_{i}({\bm r},{\bm p},t)\frac{m}{e},
\end{eqnarray}
where $e$ is the energy of a baryon, and $f_{i}({\bm r},{\bm p}, t)$ is the phase-space distribution function. 
Shown in Fig.~\ref{kaon-pot}(b) are the corresponding kaon effective masses in the two scenarios. Obviously, consistent with the tendency of kaon potential, the kaon effective mass with the chiral Lagrangian (chL) scenario is larger than that with the empirical scattering length (ESL) scenario.

\section{Results and Discussions}\label{Results and Discussions}
Now, we present the results of Au + Au collisions with a centrality of 0-40\% at $\sqrt{s_{\rm NN}}=2.4$~GeV that has been carried out by the HADES Collaboration at GSI~\cite{HADES18,HADES19}. Because we have not yet incorporated the $K^{-}$ production in our model, and also we note the $K^{0}$ involves in a transition into $K_{S}^{0}$, and the latter could decay into $\pi^{+}+\pi^{-}$, thus complicating the study, so we focus only on the $K^{+}$ production in this study. The corresponding data on $K^{+}$ can be found in Refs.~\cite{Song21,HADES18}. Before we examine the kaon production from specific reaction channels, we first show the densities reached in the reaction to have a global picture on the reaction dynamics. Shown in Fig.~\ref{den} are the densities reached in a central collision sphere with a radius of 2~fm as a function of time. It is seen that the attainable densities in the reaction are approximate 2.56$-$2.58 times the saturation density for both cases. This indicates that the reaction dynamics are not changed essentially by the kaon potential scenaros and/or the corresponding medium modification of kaon mass due to the fact that the kaon yields in the reaction are relatively small, i.e., about 0.03 per event on average~\cite{HADES18}. However, the differences of compression densities between different cases are still visible as shown in the inset. As to the resulting effects on observables, we discuss later in combination with specific results.

In HICs at SIS energies, kaons are produced mostly from the baryon-baryon and pion-baryon collisions, the corresponding cross sections we adopt in this study are taken from the Refs.~\cite{Rand80,Cugn84,Cugn90}, which are also used in ART~\cite{ART} and/or AMPT~\cite{AMPT} models. Specifically, for kaons from baryon-baryon collisions, we consider the following channels,
\begin{eqnarray}
NN&\rightarrow&{NY}K^{+}, \Delta{Y}{K^{+}},\\
NR&\rightarrow&{NY}K^{+}, \Delta{Y}{K^{+}},\\
RR&\rightarrow&{NY}K^{+}, \Delta{Y}{K^{+}},
\end{eqnarray}
where $R$ denotes the $\Delta(1232)$ (simplified as $\Delta$) and $N^{*}(1440)$ which have been included in original IBUU model~\cite{IBUU1,IBUU2}, and $Y$ denotes the $\Lambda$ and $\Sigma$ with different charge states. For the pion-baryon interactions, kaons are produced through the channels
\begin{eqnarray}
\pi + N\rightarrow Y + K^{+},\\
\pi + R\rightarrow Y + K^{+}.
\end{eqnarray}
For the momentum of kaons produced from above channels, we adopt  
the momentum distribution for the kaon from a baryon-baryon collision as parameterized in Ref.~\cite{Rand80}, while for the kaon from pion-baryon collisions, we use the isotropic distributions. Moreover, considering that elastic collisions between kaons and baryons as well as pions also affect the momentum distribution of kaons at final states, we also consider these channels using a constant cross section of 10~mb for kaon-baryon elastic collisions and the cross sections of the form given in Ref.~\cite{Ko81} for kaon-pion elastic collisions, which are also same as those used in ART~\cite{ART} and/or AMPT~\cite{AMPT} models. In addition, we also consider the kaon annihilation, i.e., the channel $K^{+}+Y\rightarrow{B}+\pi$, where $B$ denotes both nucleons and resonances, and the corresponding cross sections are evaluated by the detailed balance relation from the reverse channels, i.e., $B + \pi \rightarrow Y + K^{+}$. 

Shown in Fig.~\ref{num} is the evolution of kaon number from different channels in the reaction with two scenarios. To examine medium modification effects of kaon masses on kaon production simultaneously, the results of two scenarios are shown with and without the medium modification of kaon masses (abbreviated as with mass mod. and w/o mass mod. in all figures). First, it can be seen that the elastic collisions between kaons and baryons as well as pions are dominant compared to all of inelastic channels. Second, for the kaon production, we observe that the nucleon-resonance ($NR$) and nucleon-nucleon ($NN$) channels are the first two major contributors to kaon yields regardless of which kaon potential scenario is used and whether the medium modification of kaon masses is considered. This is consistent with the previous observation in Au + Au collisions at a beam energy of 1~GeV and with a soft EoS of incompressibility $K_{0}=200$~MeV~\cite{LiBA94}, although in which the kaon potential is neglected. Third, it is observed that the medium modification of kaon masses has obvious effects on kaon production in HICs. For example, the kaon net number is obviously smaller in the reaction with the consideration of medium modification of kaon masses than that without the consideration of medium modification of kaon masses. To confirm this observation, we further examine the rapidity and tranverse mass distributions of the kaons produced at the final states in the following. 

\begin{figure}[htb]
	\includegraphics[width=\columnwidth]{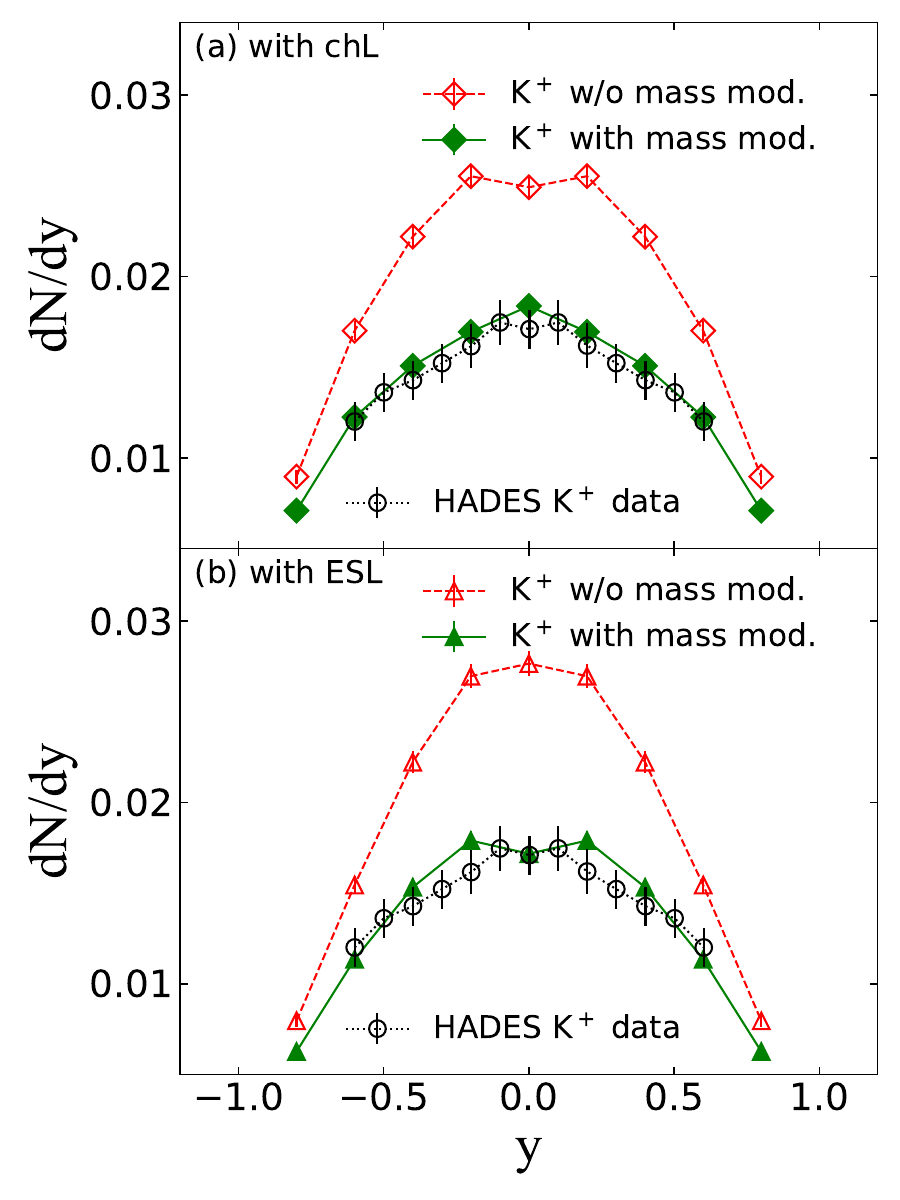}
	\caption{(Color online) Rapidity distributions of kaons produced in 0-40\% Au+Au collisions at $\sqrt{s_{\rm NN}}=2.4$~GeV in comparison with the corresponding HADES data~\cite{Song21,HADES18}.}
	\label{kaony}
\end{figure}
\begin{figure*}[thb]
	\includegraphics[width=\textwidth]{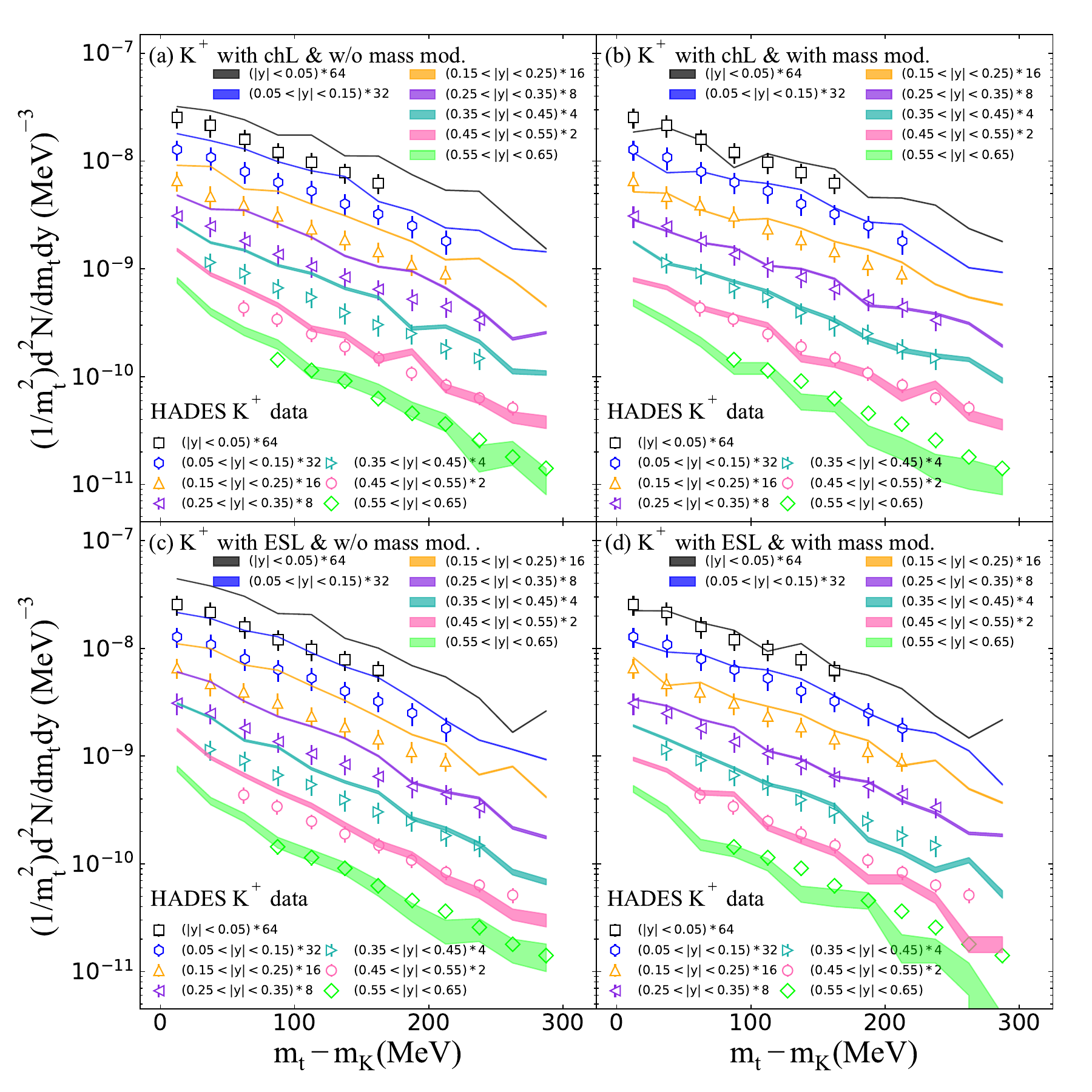}
	\caption{(Color online)Transverse mass distributions of kaons produced in 0-40\% Au+Au collisions at $\sqrt{s_{\rm NN}}=2.4$~GeV in comparison with the corresponding HADES data~\cite{Song21,HADES18}.}
	\label{kaonspr}
\end{figure*}
\begin{figure*}[thb]
	\includegraphics[width=\textwidth]{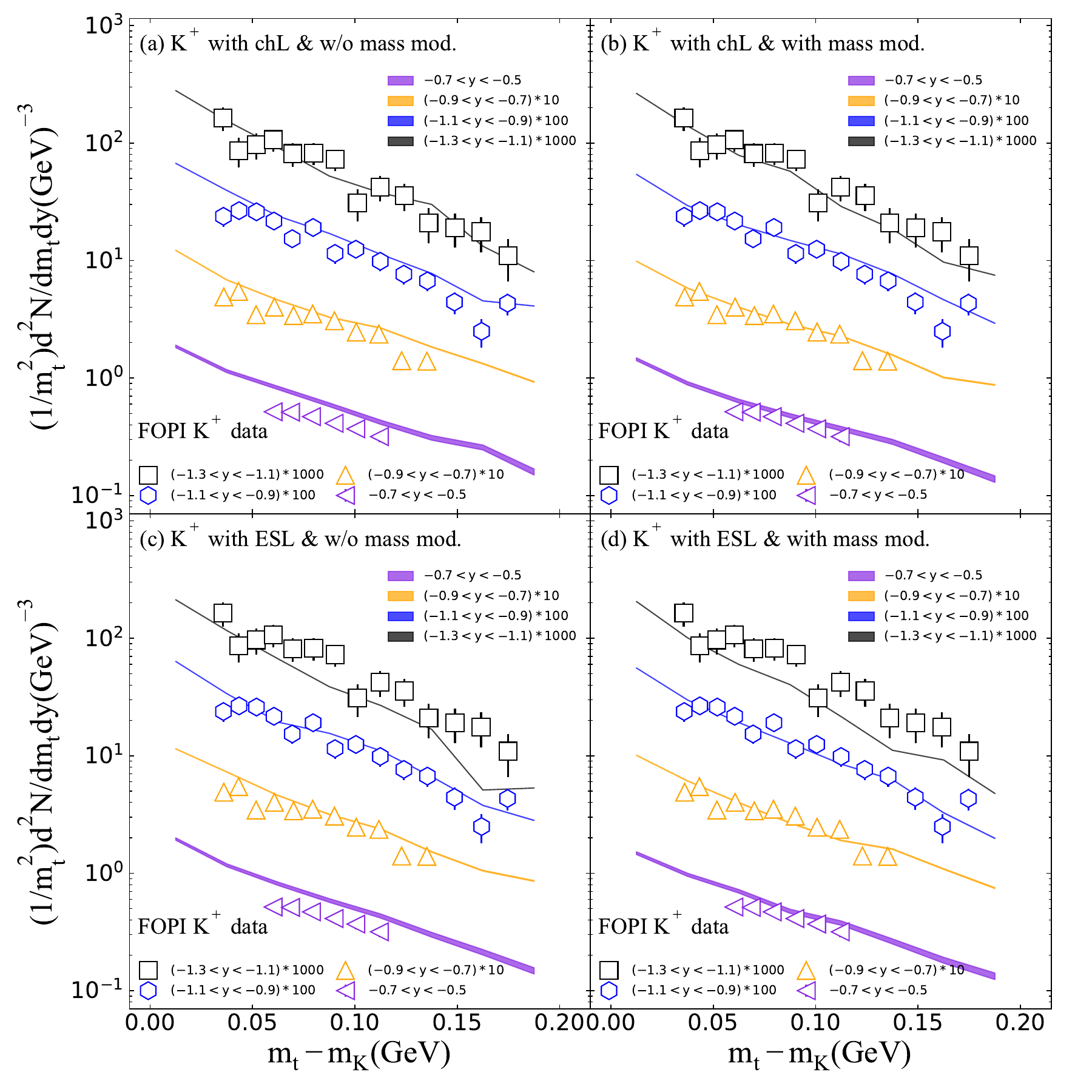}
	\caption{(Color online)Transverse mass distributions of kaons produced in Ni + Ni collisions at 1.93\textit{A} GeV with an impact parameter of 0-3.3 fm in comparison with the corresponding FOPI data~\cite{FOPI-Kaon97}.}
	\label{kaonsprNi}
\end{figure*}
\begin{figure*}[htb]
	\includegraphics[width=\textwidth]{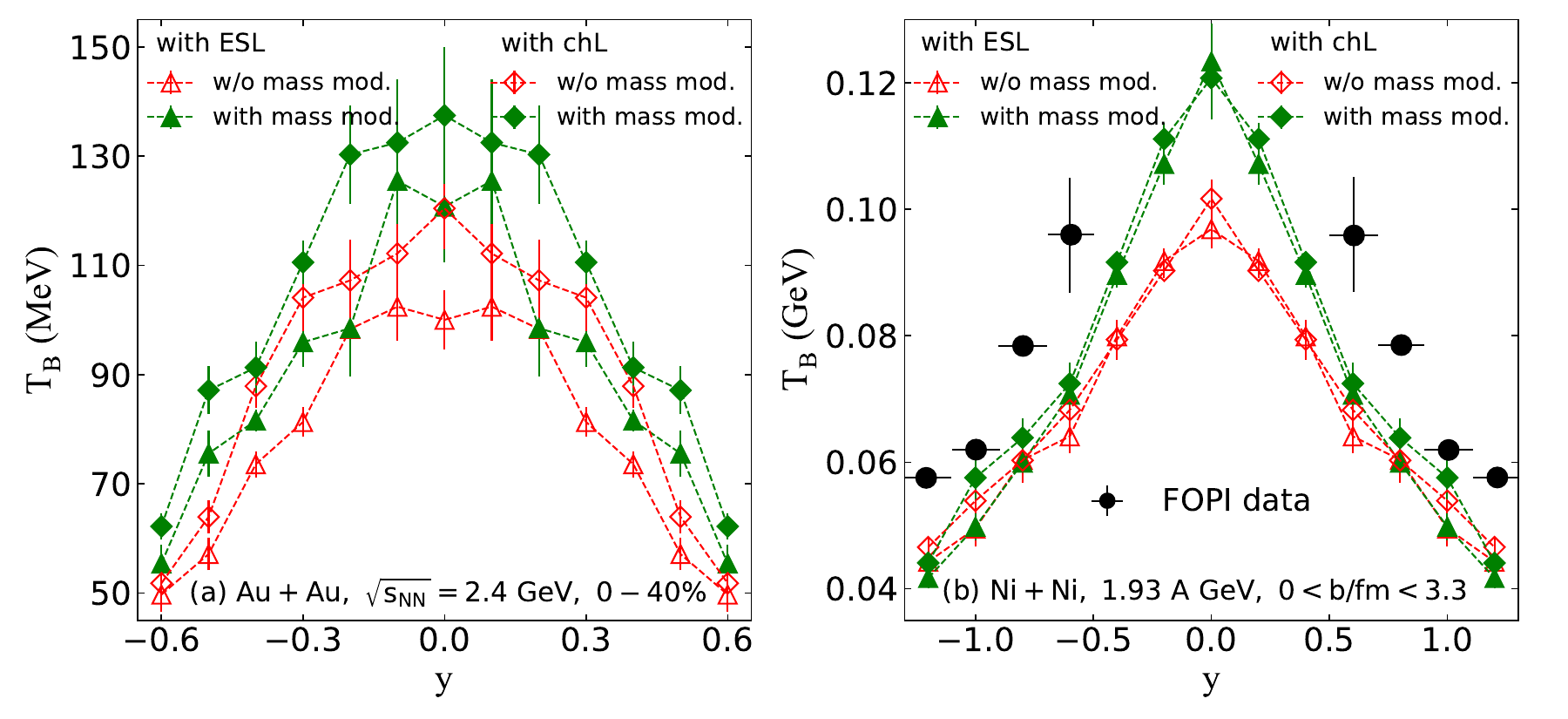}
	\caption{(Color online)Inverse slope parameter $T_{B}$ of the kaon transverse mass spectra as a function of rapidity in 0-40\% Au+Au collisions at $\sqrt{s_{\rm NN}}=2.4$~GeV (a) and in Ni + Ni collisions at 1.93\textit{A} GeV with an impact parameter of 0-3.3 fm (b). The corresponding FOPI data for Ni + Ni collisions are taken from Ref.~\cite{FOPI-Kaon97} for comparisons.}
	\label{slope}
\end{figure*}
\begin{table*}[hbt]
	\caption{\label{table}
		The calculated $\Delta{N}$, $A$, and $T_{B}$ at the midrapidity of $y=0$ for the Au + Au collisions at $\sqrt{s_{\rm NN}}=2.4$~GeV and Ni + Ni collisions at 1.93\textit{A} GeV in two kaon potential scenarios with and without the medium modification of kaon masses.}
	\begin{ruledtabular}
		\begin{tabular}{ccccccccccccc}
			\textrm{${\rm Reaction}$}&
			\textrm{$\rm Case$}&
			\textrm{$\rm \Delta{\it N}|_{y=0}$\footnote{Note that $\Delta{N}$|$_{y=0}$ denotes the number of kaons within a rapidity range with a center of $y=0$ and an interval of 0.1 for the Au + Au reactions and 0.2 for the Ni + Ni reactions. The same settings consistent with the corresponding HADES~\cite{Song21,HADES18} and FOPI~\cite{FOPI-Kaon97} experiments are also used for calculating the rapidity-dependent $A$ and $T_{B}$ at $y=0$, “w/o” stands for “without”.}}&
			\textrm{$A|_{\rm y=0}~(\rm MeV^{-3})$}&
			\textrm{$T_{B}|_{\rm y=0}~(\rm MeV)$}\\
			\colrule
			Au+Au      & ESL~w/o~mass~mod.      & $0.0278\pm0.00002$   & $8.25\times10^{-10}\pm3.09\times10^{-11}$       & $100.03\pm5.4$\\  
			& ESL~with~mass~mod.     & $0.0174\pm0.00003$   & $4.30\times10^{-10}\pm2.40\times10^{-11}$       & $120.76\pm10.1$\\
			& chL~w/o~mass~mod.      & $0.0249\pm0.00002$   & $5.95\times10^{-10}\pm2.46\times10^{-11}$       & $120.44\pm7.4$\\
			& chL~with~mass~mod.     & $0.0176\pm0.00003$   & $3.59\times10^{-10}\pm2.77\times10^{-11}$       & $137.46\pm12.5$\\             
			\hline
			Ni+Ni      & ESL~w/o~mass~mod.      & $0.094\pm0.00015$    & $2.84\times10^{-9}\pm6.25\times10^{-11}$                                   & $97\pm3.1$\\  
			& ESL~with~mass~mod.     & $0.074\pm0.00012$    & $1.73\times10^{-9}\pm5.54\times10^{-11}$                                   & $123\pm5.9$\\
			& chL~w/o~mass~mod.      & $0.095\pm0.00015$    & $2.64\times10^{-9}\pm5.34\times10^{-11}$                                   & $101\pm2.9$\\
			& chL~with~mass~mod.     & $0.071\pm0.00011$    & $1.63\times10^{-9}\pm5.86\times10^{-11}$                                   & $121\pm6.5$\\ 
		\end{tabular} 
	\end{ruledtabular}
\end{table*}

Shown in Fig.~\ref{kaony} are the rapidity distributions of final state kaons produced in the reaction with and without the medium modification of kaon masses, and also in comparison with the corresponding HADES data. Indeed, compared to the case without the medium modification of kaon masses, the kaon rapidity distributions are suppressed significantly by the medium modification of kaon masses, and also in reasonable agreement with the HADES data regardless of which scenario is used. The reasons are naturally due to the medium effect on kaon mass as shown in Fig.~\ref{kaon-pot}(b) that increases the threshold of kaon production in HICs. Actually, it is not limited to the kaon rapidity distributions, 
the kaon transverse mass ($m_{t}=\sqrt{m^{2}_{K}+p^{2}_{t}}$) spectra at different rapidity bins as shown in Fig.~\ref{kaonspr} are also found to be suppressed significantly by the medium modification of kaon masses, and again in reasonable agreement with the HADES data regardless of which scenario is used. However, we cannot easily observe the obvious differences between the two scenarios from both the kaon rapidity distributions and transverse mass spectra. Actually, as shown in the inset of Fig.~\ref{den}, the differences in compression stage between the two scenarios are visible. For example, without the consideration of medium modification of kaon mass, one can observe that the attainable densities of the reaction in compression stage is larger for the empirical scattering length scenario than that for the chiral Lagrangian scenario. This naturally leads one to expect more production of the kaons in the reaction with the empirical scattering length scenario. Also, it has been pointed out in some of the literature, e.g, in Refs.~\cite{Song21,Song22}, the inverse slope parameter $T_{B}$ of kaon transverse mass spectra is sensitive to the medium effects felt by kaons, this naturally leads us to look at the inverse slope parameter of kaon transverse mass spectra in the following.

\begin{figure*}[htb]
	\includegraphics[width=\textwidth]{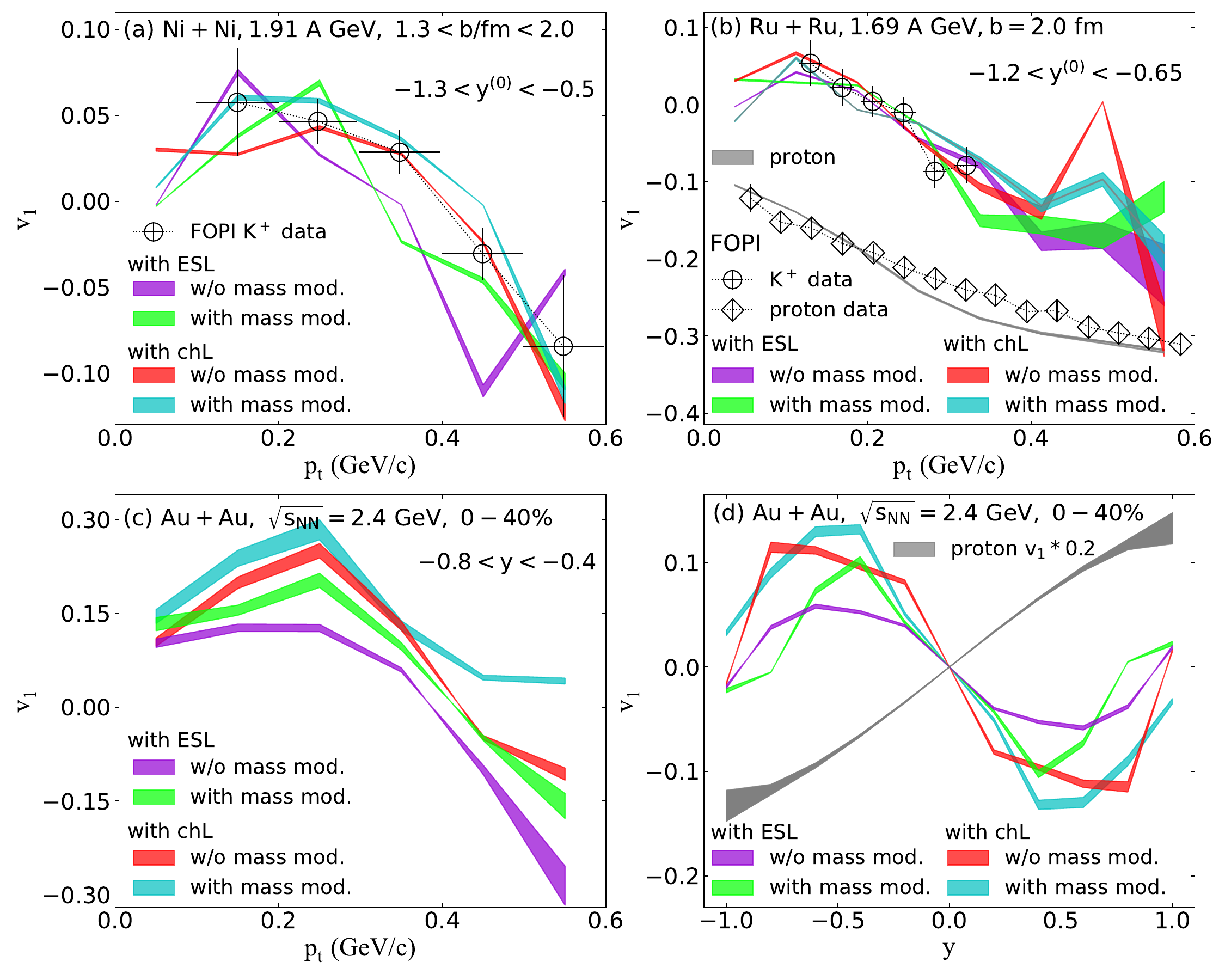}
	\caption{Transverse momentum-dependent directed flows of kaons in Ni + Ni collisions at 1.91\textit{A} GeV (a) and in Ru + Ru collisions at 1.69\textit{A} GeV (b) in comparison with the corresponding FOPI data~\cite{FOPI-Kaon00,FOPI-Kaon14,Metag17}. The directed flows of kaons in 0-40\% HADES Au+Au collisions at $\sqrt{s_{\rm NN}}=2.4$~GeV as a function of transverse momentum (c) and rapidity (d).}
	\label{flow}
\end{figure*}
\begin{figure}[thb]
	\includegraphics[width=\columnwidth]{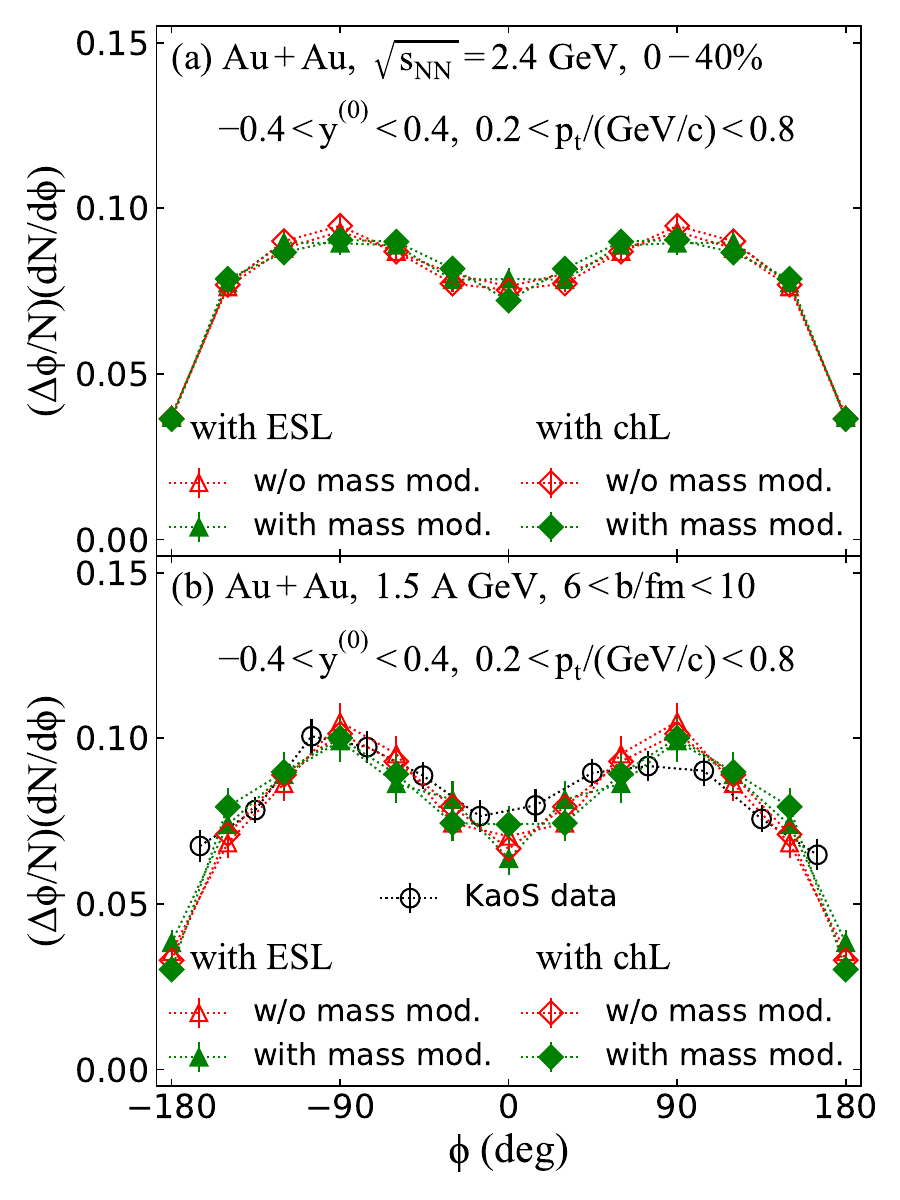}
	\caption{Azimuthal distributions of kaons in 0-40\% HADES Au+Au collisions at $\sqrt{s_{\rm NN}}=2.4$~GeV (a) and in KaoS Au + Au collisions at 1.5\textit{A} GeV with the impact parameter from 6 to 10 fm in comparison with the corresponding KaoS data~\cite{AB05}.}
	\label{ang}
\end{figure}

The inverse slope parameter $T_{B}$ is defined in the following expression~\cite{Song21,FOPI-Kaon97,FOPI-Kaon07},
\begin{equation}\label{inverse-slope}
\frac{1}{m^{2}_{t}}\frac{d^{2}N}{d(m_{t}-m_{K})dy}=A\cdot{\rm exp}\frac{-(m_{t}-m_{K})}{T_{B}},
\end{equation}
where both the $T_{B}$ and  integration constant $A$ depend on the rapidity $y$. 
Shown in Fig.~\ref{slope}(a) is the inverse slope parameter $T_{B}$ of kaon transverse mass spectra as a function of rapidity in 0-40\% Au+Au collisions at $\sqrt{s_{\rm NN}}=2.4$~GeV. It is seen that the rapidity-dependent inverse slope parameter $T_{B}$ of kaon transverse mass spectra is affected considerably by both the kaon potential and medium modification of kaon masses. To help understand this observation, we also list the specific values of $\Delta{N}$, $A$, and $T_{B}$ at the midrapidity of $y=0$ in Table~\ref{table}, where $\Delta{N}$ denotes the number of kaons within a rapidity interval of 0.1 around a center of $y=0$ as adopted in the HADES experiment as shown in Fig.~\ref{kaonspr}. Indeed, as expected, without the consideration of medium modification of kaon masses, a larger compression density reached in the reaction with the empirical scattering length scenario causes a more production of kaons (e.g., a larger value of $\Delta{N}$ at $y=0$), also due to the negative correlation between $A$ and $T_{B}$~\cite{FOPI-Kaon97}, and thus results in a smaller inverse slope parameter $T_{B}$ compared to the case with the chiral Lagrangian scenario. Moreover, for a certain kaon potential, since the medium modification of kaon masses could increase the threshold of kaon production in HICs and thus reduce the kaon production, this naturally leads to a larger inverse slope parameter $T_{B}$ compared to the case without the medium modification of kaon masses. Certainly, for the chiral Lagrangian scenario, the maximum compression seems a very tiny larger in the reaction with than without the consideration of medium modification of kaon masses. 
This is because the medium modification of kaon masses reduces the kaon production in compression stage, and thus correspondingly reduces the probability of kaons repelling nucleons away from the compression region. Naturally, this dynamic competing process could lead to a very tiny larger compression in the reaction for the case with the medium modification of kaon masses.

Fig.~\ref{kaonsprNi} and Fig.~\ref{slope}(b) show the kaon transverse mass spectra and inverse slope parameter $T_{B}$ in Ni + Ni collisions at 1.93\textit{A} GeV with an impact parameter of 0-3.3 fm in comparison with the corresponding FOPI data~\cite{FOPI-Kaon97}.The specific values of $\Delta{N}$, $A$, and $T_{B}$ at the midrapidity of $y=0$ for this reaction are also listed in Table~\ref{table}, in which the rapidity interval of 0.2 around the center of $y=0$ is adopted as in the FOPI experiment. It is seen that the differences of transverse mass spectra between four cases are not obvious, and it seems they can all fit the FOPI data fairly. However, as shown in Fig.~\ref{slope}(b), the simulated inverse slope parameter $T_{B}$ with the consideration of medium modification of kaon masses is evidently more closer to the FOPI data. Certainly, unlike the observations in HADES Au+Au collisions at $\sqrt{s_{\rm NN}}=2.4$~GeV, we indeed cannot distinguish the two scenarios from both the kaon transverse mass spectra and inverse slope parameter in Ni + Ni collisions at 1.93\textit{A} GeV regardless of the medium modification of kaon masses is considered or not.
Anyway, we can confirm from these observations that the medium modification of kaon masses plays a vital role in kaon production in HICs and also is unavoidable for the successful interpretation of both the HADES and FOPI kaon data. Predictably, measurements of the inverse slope parameter $T_{B}$ of kaon transverse mass spectra in HADES Au + Au collisions would be great benefit to detection of both the kaon potential and the corresponding medium effects on kaon masses.

So far, one might wonder how the collective flows are affected by the kaon potential scenarios and the corresponding medium modification of kaon masses, since it has been pointed out that the directed and/or transverse flows of kaons are the most promising probes of kaon potentials in HICs at both Alternating Gradient Synchrotron~\cite{Pal00} and SIS~\cite{LiGQ95} energies. Therefore, we present the directed flows of kaons as a function of the rapidity as shown in Fig.~\ref{flow}(d). For comparisons, we also show the corresponding flows of protons scaled by a factor of 0.2 in the same reaction. Because the proton flows are almost unaffected by the kaon production, we therefore present only one curve for the proton directed flows. Obviously, consistent with the observations in Refs.~\cite{Pal00,LiGQ95}, the directed flows of kaons are in the opposite direction from those of protons, i.e., the appearance of an antiflow with respect to protons. Moreover, we can also observe that the directed flows of kaons with the chiral Lagrangian scenario are larger in amplitude than those with the empirical scattering length scenario regardless of the medium modification of kaon masses is considered or not. To understand this observation, we examine the strength of specific kaon potoential in the two scenarios as shown in Fig.~\ref{kaon-pot}(a). It can be seen that the strength of  chiral Lagrangian kaon potential is larger than that of empirical scattering length kaon potential, it naturally causes a stronger repulsive effect on the kaons, and thus making the amplitude of directed flows larger. Also due to this reason, we can observe similar features in the transverse momentum $p_{t}=\sqrt{p^{2}_{x}+p^{2}_{y}}$ -dependent directed flows as shown in Fig.~\ref{flow}(c). 
Also, to verify the validity of our simulations, we show the results of simulations of Ni + Ni collisions at 1.91\textit{A} GeV and Ru + Ru collisions at 1.69\textit{A} GeV as shown in Fig.~\ref{flow}(a) and Fig.~\ref{flow}(b), respectively. Shown in Fig.~\ref{flow}(b) also includes the corresponding proton flows. To compare with the corresponding FOPI data~\cite{FOPI-Kaon00,FOPI-Kaon14,Metag17}, we use the same settings, e.g., $y^{(0)}=(y/y_{\rm beam})_{\rm c.m.}$, and the specific values of settings can be found in Fig.~\ref{flow}(a) and Fig.~\ref{flow}(b). It is seen that the directed flows of both protons and kaons are reasonably in agreement with the corresponding data. However, it seems hard to distinguish effects of different kaon potential scenarios and the corresponding medium modification of kaon masses from these FOPI data. In addition, we also present the azimuthal distributions of kaons in Au + Au collisions at 1.5\textit{A} GeV with the impact parameter from 6 to 10 fm as shown in Fig.~\ref{ang}(b), and also in comparison with the corresponding KaoS data~\cite{AB05}. It is seen that our simulations are also reasonably in agreement with the corresponding data. Shown in Fig.~\ref{ang}(a) are the azimuthal distributions of kaons in HADES Au + Au collisions, in which the same rapidity settings as in Fig.~\ref{ang}(b) are adopted. Unfortunately, we do not find obvious differences between the two scenarios from the azimuthal distributions of kaons in both KaoS Au + Au collisions at 1.5\textit{A} GeV and HADES Au + Au collisions at $\sqrt{s_{\rm NN}}=2.4$~GeV, as shown in Fig.~\ref{ang}(b) and Fig.~\ref{ang}(a), respectively. From these observations especially the directed flows of kaons as shown in Fig.~\ref{flow}, we can observe that the kaon potential scenario indeed has regular effect on the flows of kaons in HADES Au + Au collision at $\sqrt{s_{\rm NN}}=2.4$~GeV and the medium modification of kaon masses also slightly affects the kaon directed flows in this reaction, however, this regular effect does not seem to be obvious for lighter reaction systems at higher beam energies, e.g., Ru + Ru at 1.69\textit{A} GeV. Therefore, measurements of the kaon directed flows in HADES Au + Au collisions might also be useful to detect both the kaon potential and the corresponding medium effects on kaon masses.

\section{Summary}\label{Summary}
In summary, we have studied the kaon production in HICs at SIS energies within an IBUU transport model, in which the kaon production, annihilation and elastic channels as well as two kaon potential scenarios and the corresponding medium modification of kaon masses are considered.
It is shown that the medium modification of kaon masses plays a vital role in studying the kaon productions in HICs at SIS energies, and is also unavoidable for the successful interpretation of the HADES kaon rapidity distributions and transverse mass spectra data. Moreover, the rapidity-dependent inverse slope parameter $T_{B}$ of the kaon transverse mass spectra in HADES Au + Au collisions is shown to be affected considerably by both the kaon potential and medium modification of kaon masses. The kaon directed flows in HADES Au + Au collisions are also shown to be affected by both the kaon potential and medium modification of kaon masses. However, in the reactions involved in FOPI and/or KaoS experiments, some of these regular effects do not seem to be obvious and appear to be the reaction system and/or beam energy dependent. Therefore, we suggest to measure the inverse slope parameter $T_{B}$ of kaon transverse mass spectra and the kaon directed flows in HADES Au + Au collisions, and thus benefit to the detection of the kaon potential and the corresponding medium effects on kaon masses.

%\begin{acknowledgments}
This work is supported by the National Natural Science Foundation of China under Grant No.11965008 and No.11405128 and Guizhou Science and Technology Foundation under Grant No.[2020]1Y034, Guizhou Provincial Science and Technology Plan Project (Guizhou Provincial Science Cooperation Platform Talents [2019]), and the PhD-funded project of Guizhou Normal university (Grant No.GZNUD[2018]11).
%\end{acknowledgments}


\begin{thebibliography}{99}
%
\bibitem{Hor14} C. J. Horowitz, E. F. Brown, Y. Kim, W. G. Lynch, R. Michaels, A. Ono, J. Piekarewicz, M. B. Tsang, H. H. Wolter, J. Phys. G: Nucl. Part. Phys. \textbf{41}, 093001 (2014).
\bibitem{Heb15} K. Hebeler, J. D. Holt, J. Men\'{e}ndez, A. Schwenk, Annu. Rev. Nucl. Part. Sci. \textbf{65}, 457 (2015).
\bibitem{Lat16} J. M. Lattimer, M. Prakash, Phys. Rep. \textbf{621}, 127 (2016).
\bibitem{Bal16} M. Baldo, G. F. Burgio, Prog. Part. Nucl. Phys. \textbf{91}, 203 (2016).
\bibitem{Oer17} M. Oertel, M. Hempel, T. Kl\"ahn, S. Typel, Rev. Mod. Phys. \textbf{89}, 015007 (2017).
\bibitem{Hana17} M. Hanauske, K. Takami, L. Bovard, L. Rezzolla, J. A. Font, F. Galeazzi, and H. St\"ocker, Phys. Rev. D \textbf{96}, 043004 (2017).
\bibitem{Kap86} D. B. Kaplan, A. E. Nelson, Phys. Lett. B \textbf{175}, 57 (1986).
\bibitem{Aich85} J. Aichelin, C. M. Ko, Phys. Rev. Lett. \textbf{55}, 2661 (1985).
\bibitem{LiGQ94} G. Q. Li, C. M. Ko, X. S. Fang, Phys. Lett. B \textbf{329}, 149 (1994).
\bibitem{LiBA94} B. A. Li, Phys. Rev. C \textbf{50}, 2144 (1994).
\bibitem{Song99} G. Song, B. A. Li, C. M. Ko, Nucl. Phys. A \textbf{646}, 481 (1999).
\bibitem{Fuch01} C. Fuchs, A. Faessler, E. Zabrodin, Y. M. Zheng, Phys. Rev. Lett. \textbf{86}, 1974 (2001).
\bibitem{Sturm01} C. Sturm, I. B\"ottcher, M. Debowski $et~al.$, Phys. Rev. Lett. \textbf{86}, 39 (2001).
\bibitem{Hart06} Ch. Hartnack, H. Oeschler, J. Aichelin, Phys. Rev. Lett. \textbf{96}, 012302 (2006).
\bibitem{Hart12} C. Hartnack, H. Oeschler, Y. Leifels, E. L. Bratkovskaya, J. Aichelin, Phys. Rep. \textbf{510}, 119 (2012).
\bibitem{Scha94} J. Schaffner, A. Gal, I. N. Mishustin, H. St\"ocker, W. Greiner, Phys. Lett. B \textbf{334}, 268 (1994).
\bibitem{Brown94} G. E. Brown, C. H. Lee, M. Rho, and V. Thorsson, Nucl. Phys. A \textbf{567}, 937 (1994).
\bibitem{LiGQ97} G. Q. Li, C. H. Lee, and G. E. Brown, Nucl. Phys. A \textbf{625}, 372 (1997).
\bibitem{LiGQ97b} G. Q. Li, C. H. Lee, and G. E. Brown, Phys. Rev. Lett. \textbf{79}, 5214 (1997).
\bibitem{Cassing97} W. Cassing, E. L. Bratkovskaya, U. Mosel, S. Teis, and A. Sibirtsev, Nucl. Phys. A \textbf{614}, 415 (1997).
\bibitem{AB05} A. B. Larionov, U. Mosel, Phys. Rev. C \textbf{72}, 014901 (2005).
\bibitem{Gasik16} P. Gasik, K. Piasecki, N. Herrmann $et~al.$, Eur. Phys. J. A \textbf{52}, 177 (2016).
\bibitem{Song21} T. Song, L. Tolos, J. Wirth, J. Aichelin, and E. Bratkovskaya, Phys. Rev. C \textbf{103}, 044901 (2021).
\bibitem{Song22} T. Song, J. Aichelin, E. Bratkovskaya, Phys. Rev. C \textbf{106}, 024903 (2022).
\bibitem{KaoS94} D. Mi$\acute{s}$kowiec, W. Ahner, R. Barth $et~al.$, Phys. Rev. Lett. \textbf{72}, 3650 (1994).
\bibitem{KaoS97} R. Barth, P. Senger, W. Ahner $et~al.$ (KaoS Collaboration), Phys. Rev. Lett. \textbf{78}, 4007 (1997).
\bibitem{KaoS98} Y. Shin, W. Ahner, R. Barth $et~al.$ (KaoS Collaboration), Phys. Rev. Lett. \textbf{81}, 1576 (1998).
\bibitem{KaoS05} F. Uhlig, A. F\"orster, I. B\"ottcher $et~al.$ (KaoS Collaboration), Phys. Rev. Lett. \textbf{95}, 012301 (2005).
\bibitem{KaoS07} A. F\"orster, F. Uhlig, I. B\"ottcher $et~al.$ (KaoS Collaboration), Phys. Rev. C \textbf{75}, 024906 (2007).
\bibitem{FOPI-Kaon97}D. Best, N. Herrmann, B. Hong $et~al.$, Nucl. Phys. A \textbf{625}, 307 (1997).
\bibitem{FOPI-Kaon00} P. Crochet, N. Herrmann, K. Wi$\acute{s}$niewski $et~al.$ (FOPI Collaboration), Phys. Lett. B \textbf{486}, 6 (2000).
\bibitem{FOPI-Kaon07} M. Merschmeyer, X. Lopez, N. Bastid $et~al.$ (FOPI Collaboration), Phys. Rev. C \textbf{76}, 024906 (2007).
\bibitem{FOPI-Kaon09} M. L. Benabderrahmane, N. Herrmann, K. Wi$\acute{s}$niewski $et~al.$ (FOPI Collaboration), Phys. Rev. Lett. \textbf{102}, 182501 (2009).
\bibitem{FOPI-Kaon14} V. Zinyuk, T. I. Kang, Y. Leifels $et~al.$ (FOPI Collaboration), Phys. Rev. C \textbf{90}, 025210 (2014).
\bibitem{HADES09} G. Agakishiev, A. Balanda, B. Bannier $et~al.$ (HADES Collaboration), Phys. Rev. C \textbf{80}, 025209 (2009).
\bibitem{HADES10} G. Agakishiev, A. Balanda, B. Bannier $et~al.$ (HADES Collaboration), Phys. Rev. C \textbf{82}, 044907 (2010).
\bibitem{HADES11} G. Agakishiev, A. Balanda, B. Bannier $et~al.$ (HADES Collaboration),  Eur. Phys. J. A \textbf{47}, 21 (2011).
\bibitem{HADES14} G. Agakishiev, O. Arnold, D. Belver $et~al.$ (HADES Collaboration), Phys. Rev. C \textbf{90}, 054906 (2014).
\bibitem{HADES18} J. Adamczewski-Musch, O. Arnold, C. Behnke $et~al.$ (HADES Collaboration), Phys. Lett. B \textbf{778}, 403 (2018). 
\bibitem{HADES19} J. Adamczewski-Musch, O. Arnold, C. Behnke $et~al.$ (HADES Collaboration), Phys. Lett. B \textbf{793}, 457 (2019). 
\bibitem{HADES20}J. Adamczewski-Musch O. Arnold, C. Behnke {\it et al}. (HADES Collaboration), Phys. Rev. Lett. \textbf{125}, 262301 (2020).
\bibitem{Pal00} S. Pal, C. M. Ko, Z. W. Lin, and B. Zhang, Phys. Rev. C \textbf{62}, 061903 (2000).
\bibitem{Metag17} V. Metag, M. Nanova, and E.Ya. Paryev, Prog. Part. Nucl. Phys. \textbf{97}, 199 (2017).
\bibitem{LiGQ95} G. Q. Li, C. M. Ko, B. A. Li, Phys. Rev. Lett. \textbf{74}, 235 (1995).
\bibitem{LiBA96} B. A. Li, C. M. Ko, Phys. Rev. C \textbf{54}, 3283 (1996).
\bibitem{LiBA99} B. A. Li, B. Zhang, A. T. Sustich, and C. M. Ko, Phys. Rev. C \textbf{60}, 034902 (1999).
\bibitem{ECT} M. Szala, Light nuclei formation in heavy ion collisions measured with HADES, report, \url{https://indico.ectstar.eu/event/52/contributions}
\bibitem{Wei23} H. Du, G. F. Wei, G. C. Yong, Phys. Lett. B \textbf{839}, 137823 (2023).
\bibitem{IBUU1} B. A. Li, C. B. Das, S. Das Gupta, C. Gale, Phys. Rev. C \textbf{69}, 011603(R) (2004).
\bibitem{IBUU2} B. A. Li, C. B. Das, S. Das Gupta, C. Gale, Nucl. Phys. A \textbf{735}, 563 (2004).
\bibitem{Xu10} C. Xu, B. A. Li, Phys. Rev. C \textbf{81}, 044603 (2010).%H1:29
\bibitem{Chen14} L. W. Chen, C. M. Ko, B. A. Li, C. Xu, J. Xu, Eur. Phys. J. A \textbf{50}, 29 (2014).%H1:30
\bibitem{Wei20} G. F. Wei, C. Xu, W. Xie, Q. J. Zhi, S. G. Chen, Z. W. Long, Phys. Rev. C \textbf{102}, 024614 (2020).
\bibitem{Cozma18} M. D. Cozma, Eur. Phys. J. A \textbf{54}, 40 (2018).
\bibitem{Wang18} R. Wang, L. W. Chen, Y. Zhou, Phys. Rev. C \textbf{98}, 054618 (2018).%26%d:27,35.5±3.2
\bibitem{Ess21} R. Essick, I. Tews, P. Landry, A. Schwenk, Phys. Rev. Lett. \textbf{127}, 192701 (2021).
\bibitem{Ree21}B. T. Reed, F. J. Fattoyev, C. J. Horowitz, J. Piekarewicz, Phys. Rev. Lett. \textbf{126}, 172503 (2021).
\bibitem{Est21} J. Estee, W. G. Lynch, C. Y. Tsang $et~al.$ ($\mathrm{S}\ensuremath{\pi}\mathrm{RIT}$ Collaboration), Phys. Rev. Lett. \textbf{126}, 162701 (2021).%SπRIT2021
\bibitem{Wang13} N. Wang, L. Ou, M. Liu, Phys. Rev. C \textbf{87}, 034327 (2013).%d:29,K0=230±30，25.5±1
\bibitem{Wei22} X. Huang, G. F. Wei, Q. J. Zhi, Y. C. Yang, Z. W. Long, Phys. Rev. C \textbf{106}, 014604 (2022).%4
\bibitem{Xu15} J. Xu, L. W. Chen, B. A. Li, Phys. Rev. C \textbf{91}, 014611 (2015).%H1-32,k0=230
\bibitem{LiBA95} B. A. Li, C. M. Ko, Phys. Rev. C \textbf{52}, 2037 (1995).
\bibitem{ART} B. A. Li, A. T. Sustich, B. Zhang, and C. M. Ko, Int. J. Mor. Phys. E \textbf{10}, 267 (2001).
%\bibitem{GY03} Y. Gao, X. G. Li, D. J. Jia, Chin. Phys. C\textbf{27}, 995 (2003) (In Chinese).
%\bibitem{Ko87} C. M. Ko, Q. Li, R. C. Wang, Phys. Rev. Lett.\textbf{59}, 1084 (1987).
\bibitem{Rand80} J. Randrup, C. M. Ko, Nucl. Phys. A \textbf{343}, 519 (1980).
\bibitem{Cugn84} J. Cugnon, R. M. Lombard, Nucl. Phys. A \textbf{422}, 635 (1984).
\bibitem{Cugn90} J. Cugnon, P. Deneye, J. Vandermeulen, Phys. Rev. C \textbf{41}, 1701 (1990).
\bibitem{AMPT} Z. W. Lin, C. M. Ko, B. A. Li, B. Zhang, and S. Pal, Phys. Rev. C \textbf{72}, 064901 (2005).
\bibitem{Ko81} C. M. Ko, Phys. Rev. C \textbf{23}, 2760 (1981).

\end{thebibliography}
\end{document}